\documentstyle[12pt,qqa4lart]{article}
\input tcilatex

\QQQ{Language}{
American English
}

\begin{document}

\author{Lu-Ming Duan and Guan-Can Guo\thanks{%
Electronic address: gcguo@sunlx06.nsc.ustc.edu.cn} \\
Dept. of Phys. and Nonlinear Science Center,\\
University of Science and Technology of China,\\
Hefei 230026, People's Republic of China}
\title{Prevention of dissipation with two particles}
\date{}
\maketitle

\begin{abstract}
\baselineskip 24ptAn error prevention procedure based on two-particle
encoding is proposed for protecting an arbitrary unknown quantum state from
dissipation, such as phase damping and amplitude damping. The schemes, which
exhibits manifestation of the quantum Zeno effect, is effective whether
quantum bits are decohered independently or cooperatively. We derive the
working condition of the scheme and argue that this procedure has feasible
practical implementation.\\

{\bf PACS numbers:} 03.65.Bz, 89.70.+c, 42.50.Dv
\end{abstract}

\newpage\baselineskip 24ptDecoherence and loss limit the practicality of
quantum cryptography and computing [1,2]. To circumvent this difficulty,
successful quantum error correction techniques have been developed [3-15].
In these schemes, some redundancy is introduced to protect the original
information. If mere a few qubits are subject to errors in a noisy
environment, these errors can be corrected by detecting the error syndromes,
without violation of the original information. On the one hand, quantum
error correction holds the promise of reliable storage, processing, and
transfer of quantum information; on the other hand, it is rather costly of
computing resources. For correcting single-qubit errors, one needs at least
five qubits to encode one qubit information [7,8]. The encoding, decoding,
and detection of error syndromes are also involved.

Compared with the conventional error correction schemes, alternate
decoherence-reducing strategies based on the quantum Zeno effect are more
efficient to implement, though they may fail for some models of
decoherence[16,17]. The use of the Zeno effect for correcting or for
preventing errors in quantum computers was first suggested by Zurek [18],
and it is a part of a scheme considered by Barenco {\it et al. }[19].
Recently, Vaidman {\it et al.} proposed an error prevention scheme of this
kind based on four-particle encoding [20]. It is shown there that four is
the minimal number of particles required for prevention of general errors.
Nevertheless, if one has more knowledge about the errors, simpler codes can
be found. It is discovered that two qubits are enough for preventing pure
dephasing due to phase damping [21,20], or for preventing pure loss due to
balanced amplitude damping to a reservoir at absolute zero temperature
[22,23]. In the error correction schemes mentioned above, different qubits
are assumed to be decohered independently. This is an ideal case. As another
ideal case, if two adjacent qubits are decohered completely collectively,
there exists an alternate error prevention scheme based on two-particle
encoding which utilizes the coherence-preserving states of qubit-pairs [24].

In this paper, we propose an error prevention scheme based on two-particle
encoding for reducing a large range of decoherence. The scheme exhibits
manifestation of the quantum Zeno effect. Compared with the known error
prevention schemes, this scheme has two favorable features. First, it covers
a large range of decoherence. The scheme is designed for preventing phase
damping and general amplitude damping. The reservoir may be at arbitrary
temperature, and the damping coefficients need not be balanced among
different qubits. Furthermore, the scheme works whether the qubits are
decohered independently or cooperatively. Second, it has a high efficiency.
Two qubits are enough for encoding one qubit. The encoding, decoding, and
error-detection in this scheme are quite simple. We also extend the scheme
for protecting states of multiple qubits. It is shown that approximately $%
L+\frac 12\log _2\left( \frac{\pi L}2\right) $ qubits are enough for
encoding $L$ qubit. The scheme costs less computing resources, so it is
easier to be implemented in practice.

First, we look at two qubits, which are subject to decoherence described by
phase damping or amplitude damping to a reservoir at arbitrary temperature .
The qubits are described by Pauli's operators $\overrightarrow{\sigma }_l$ $%
\left( l=1,2\right) $ and the reservoir is modelled by a bath of
oscillators. The bath modes coupling to the $l$ qubit are denoted by $%
a_{\omega l}$ ($\omega $ varies from $0$ to $\infty $ and $l=1,2$). Some of
the modes $a_{\omega 1}$ and $a_{\omega 2}$ are possibly the same and some
of them are different. We use the notation $\stackrel{2}{\stackunder{l=1}{%
\bigcup }}a_{\omega l}^{+}a_{\omega l}$ to indicate the joint sum of $%
a_{\omega l}^{+}a_{\omega l}$, so $\stackrel{2}{\stackunder{l=1}{\bigcup }}%
a_{\omega l}^{+}a_{\omega l}=a_{\omega 1}^{+}a_{\omega 1}+a_{\omega
2}^{+}a_{\omega 2}$ if $a_{\omega 1}$ and $a_{\omega 2}$ belong to different
modes and $\stackrel{2}{\stackunder{l=1}{\bigcup }}a_{\omega l}^{+}a_{\omega
l}=a_{\omega 1}^{+}a_{\omega 1}$ if $a_{\omega 1}$ and $a_{\omega 2}$ are
the same. With this notation, the Hamiltonian describing dissipation of the
two qubits has the following form (setting $\hbar =1$) 
\begin{equation}
\label{1}H=\omega _0\left( \sigma _1^z+\sigma _2^z\right) +\stackrel{2}{%
\stackunder{l=1}{\sum }}\int d\omega \left[ g_{\omega l}A_l\left( a_{\omega
l}^{+}+a_{\omega l}\right) \right] +\int d\omega \stackrel{2}{\stackunder{l=1%
}{\bigcup }}\left( \omega a_{\omega l}^{+}a_{\omega l}\right) ,
\end{equation}
where the coping coefficients $g_{\omega l}$ may be dependent of $\omega $
and $l$. The qubit operator $A_l$ in general is expressed as a linear
superposition of three Pauli's operators, i.e., $A_l=\lambda ^{\left(
1\right) }\sigma _l^x+\lambda ^{\left( 2\right) }\sigma _l^y+\lambda
^{\left( 3\right) }\sigma _l^z$. The ratio $\lambda ^{\left( 1\right)
}:\lambda ^{\left( 2\right) }:\lambda ^{\left( 3\right) }$ is determined by
the type of the dissipation. For example, $\lambda ^{\left( 1\right)
}=\lambda ^{\left( 2\right) }=0$ for phase damping and $\lambda ^{\left(
3\right) }=0$ for amplitude damping [25]. Phase damping induces pure
dephasing and amplitude damping induces loss and dephasing at the same time.
Many sources of decoherence in quantum computers can be described by
amplitude damping [26].

The qubit operator $A_l$ satisfies the condition $tr\left( A_l\right) =0$,
so without loss of generality , its two eigenvectors are denoted by $\left|
\pm 1\right\rangle _l$, with the eigenvalues $\pm 1$, respectively. The
physical basis vectors $\left| \pm \right\rangle _l$ are eigenstates of the
operator $\sigma _l^z$. The states $\left| \pm 1\right\rangle _l$ in general
differ from $\left| \pm \right\rangle _l$ by a single-qubit rotation
operation $R_l\left( \theta \right) $, i.e., $\left| \pm 1\right\rangle
_l=R_l\left( \theta \right) \left| \pm \right\rangle _l$, where $\theta $
depends on the type of the dissipation. The computation basis vectors $%
\left| 0\right\rangle _l$ and $\left| 1\right\rangle _l$ in this paper are
defined by 
\begin{equation}
\label{2}
\begin{array}{c}
\left| 0\right\rangle _l=\frac 1{
\sqrt{2}}\left( \left| +1\right\rangle _l+\left| -1\right\rangle _l\right) ,
\\  \\ 
\left| 1\right\rangle _l=\frac 1{\sqrt{2}}\left( \left| +1\right\rangle
_l-\left| -1\right\rangle _l\right) . 
\end{array}
\end{equation}
The are derived from $\left| \pm 1\right\rangle _l$ by a Hadamard
transformation $H_l$. The states $\left| 0\right\rangle _l$ and $\left|
1\right\rangle _l$ have the important property that they are flipped by the
qubit operator $A_l$. In general, the computation basis and the physical
basis differ by a single-qubit rotation operation. But if $\lambda ^{\left(
3\right) }=0$, i.e., for amplitude damping, these two bases reduce to the
same.

We use the two dissipative qubits to protect one qubit information. The
initial state of one qubit ( the information carrier) is generally expressed
as $\left| \Psi \left( 0\right) \right\rangle _1=c_{+}\left| +\right\rangle
_1+c_{-}\left| -\right\rangle _1$. The other qubit (the ancilla) is
prearranged in the state $\left| +\right\rangle _2$. We use the symbol $%
C_{ij}$ to denote the quantum controlled NOT (CNOT) operation in the
physical basis, where the first subscript of $C_{ij}$ refers to the control
bit and the second to the target. The input state of the information carrier
is encoded into the following state 
\begin{equation}
\label{3}\left| \Psi \left( 0\right) \right\rangle _1\otimes \left|
+\right\rangle _2\stackrel{C_{12}R_1\left( \theta \right) R_2\left( \theta
\right) H_1H_2}{\longrightarrow \longrightarrow }\left| \Psi
_{enc}\right\rangle =c_{+}\left| 01\right\rangle +c_{-}\left|
10\right\rangle . 
\end{equation}
For amplitude damping, the above joint operation $C_{12}R_1\left( \theta
\right) R_2\left( \theta \right) H_1H_2$ reduces to a simple CNOT $C_{12}$
in the physical basis. But for more general dissipation, this encoding
requires the knowledge of the noise parameters $\lambda ^{\left( 1\right)
}:\lambda ^{\left( 2\right) }:\lambda ^{\left( 3\right) }$. The encoding
space spanned by the sates $\left| 01\right\rangle $ and $\left|
10\right\rangle $ is denoted by the symbol $S_0$, which is a subspace of the
whole $2\times 2$ dimensional Hilbert space of the two qubits.

For pure amplitude damping, states in the encoding space $S_0$ are left
unchanged by the free Hamiltonian $H_0=\omega _0\left( \sigma _1^z+\sigma
_2^z\right) $ of the qubits. But for more general dissipation, the
Hamiltonian $H_0$ may map the initial encoded state out of the encoding
space. This probability should be avoided. The free-Hamiltonian-elimination
(FHE) technique is introduced to attain this goal. We apply a homogeneous
classical far-violet-detuned optical field $E$ to the two qubits. Under the
adiabatic approximation, the additional Hamiltonian describing the driving
process reads [27] 
\begin{equation}
\label{4}H_{drv}=-\stackrel{2}{\stackunder{l=1}{\sum }}\frac{2\left|
g\right| ^2\left| E\right| ^2}{\omega _{opt}-\omega _0}\sigma _l^z, 
\end{equation}
where $\omega _{opt}$ is the frequency of the optical field. By adjusting
the intensity $\left| E\right| ^2$ of the optical field, we choose the
coefficient in Eq. (4) to satisfy $\frac{2\left| g\right| ^2\left| E\right|
^2}{\omega _{opt}-\omega _0}=\omega _0$. The effect of the free Hamiltonian $%
H_0$ is thus offset by the driving field. Compared with the FHE procedure
involved in Ref. [24], this technique has the advantage that it operates
without the knowledge of the noise parameters.

Our scheme is based on the quantum Zeno effect. The protection procedure
consists of frequent tests that the two-qubit system has not left the
encoding space. Now we show that an arbitrary encoded state is indeed frozen
through these tests. The initial density operator of the reservoir is
denoted by $\rho _r\left( 0\right) $. Suppose $\left| \Psi _{ru}\left(
0\right) \right\rangle $ is a purification of $\rho _r\left( 0\right) $,
i.e., the state $\left| \Psi _{ru}\left( 0\right) \right\rangle $ satisfies $%
tr_u\left( \left| \Psi _{ru}\left( 0\right) \right\rangle \left\langle \Psi
_{ru}\left( 0\right) \right| \right) =\rho _r\left( 0\right) $, where the
symbol $u$ denotes an ancillary system. During a finite time $T_0$, we
perform $N$ times tests. In a short period of time $T_0/N$, under the
Hamiltonian (1) and (4), the encoded state (3) evolves into 
\begin{equation}
\label{5}
\begin{array}{c}
\left| \Psi \left( t\right) \right\rangle \approx \left[ 1-i\left(
H+H_{drv}\right) T_0/N\right] \left| \Psi _{enc}\right\rangle \otimes \left|
\Psi _{ru}\left( 0\right) \right\rangle  \\  
\\ 
=\left| \Psi _{enc}\right\rangle \otimes \left[ 1-\frac 1N\int d\omega 
\stackrel{2}{\stackunder{l=1}{\bigcup }}\left( i\omega T_0a_{\omega
l}^{+}a_{\omega l}\right) \right] \left| \Psi _{ru}\left( 0\right)
\right\rangle  \\  \\ 
-\frac 1N\left[ \left| 00\right\rangle \otimes \left(
c_{+}X_2+c_{-}X_1\right) \left| \Psi _{ru}\left( 0\right) \right\rangle
+\left| 11\right\rangle \otimes \left( c_{+}X_1+c_{-}X_2\right) \left| \Psi
_{ru}\left( 0\right) \right\rangle \right] ,
\end{array}
\end{equation}
where $X_l=\int d\omega \left[ ig_{\omega l}T_0\left( a_{\omega
l}^{+}+a_{\omega l}\right) \right] $. This evolution has two important
properties: First, after the evolution the amplitude of the state outside
the encoding space is of the order of $1/N$. Hence if we perform a
measurement in succession to tell us whether the two-qubit system has left
the encoding space $S_0$, the probability for getting the result ''out of $%
S_0$'' is of the order of $1/N^2$. Second, after the evolution the amplitude
of the state inside the encoding space remains the same as the initial
encoded state. Therefore, if we get the result '' in $S_0$'' in the
measurement, the encoded state is unchanged by the dissipation. The
two-qubit system after the test without a postselection of the measurement
results is in a mixed state, whose density operator is represented by $\rho
\left( T_0/N\right) $. The operator $\rho \left( T_0/N\right) $ can be
expressed as 
\begin{equation}
\label{6}\rho \left( T_0/N\right) =\stackrel{\symbol{94}}{S}\left(
T_0/N\right) \rho \left( 0\right) ,
\end{equation}
where $\rho \left( 0\right) =\left| \Psi _{enc}\right\rangle \left\langle
\Psi _{enc}\right| $ and $\stackrel{\symbol{94}}{S}\left( T_0/N\right) $ is
a superoperator. Obviously $\stackrel{\symbol{94}}{S}\left( T_0/N\right) $
has the following decomposition 
\begin{equation}
\label{7}\stackrel{\symbol{94}}{S}\left( T_0/N\right) =\stackrel{\symbol{94}%
}{I}+O\left( 1/N^2\right) ,
\end{equation}
where $\stackrel{\symbol{94}}{I}$ is the unit superoperator. After the time $%
T_0$, the final state $\rho \left( T_0\right) $ of the qubits is thus 
\begin{equation}
\label{8}\rho \left( T_0\right) =\left[ \stackrel{\symbol{94}}{S}\left(
T_0/N\right) \right] ^N\rho \left( 0\right) =\left| \Psi _{enc}\right\rangle
\left\langle \Psi _{enc}\right| +O\left( 1/N\right) .
\end{equation}
This equation suggests that the difference between the final and the initial
state of the qubits is of the order of $1/N$ and can be neglected for large $%
N$. After decoding the final state, we successfully protect one qubit
information from decoherence by the two dissipative qubits.

The remaining question is how to test that the two qubits has not left the
encoding space. The states $\left| 0\right\rangle _l$ and $\left|
1\right\rangle _l$ are orthogonal to each other, so they are eigenstates of
an observable $B_l$. We perform a quantum non-demolition (QND) measurement
of the observable $B_1+B_2$ on the two qubits. The states of the qubits lie
in the encoding space if and only if the measurement of $B_1+B_2$ yields the
result $1$. The QND measurement of $B_1+B_2$ can be done in the following
way: First prepare an ancilla qubit $3$ in the state $\left| 0\right\rangle
_3$ and then apply a joint operation $C_{13}^{^{\prime }}C_{23}^{^{\prime }}$
to the three qubits, where $C_{ij}^{^{\prime }}=R_i\left( \theta \right)
R_j\left( \theta \right) H_iH_jC_{ij}H_iH_jR_i\left( -\theta \right)
R_j\left( -\theta \right) $ is the quantum CNOT operation in the computation
basis $\left\{ \left| 01\right\rangle ,\left| 10\right\rangle \right\} $. By
testing whether the state of the ancilla qubit $3$ has been flipped, we
perform a QND measurement of the observable $B_1+B_2$.

Eq. (8) is approximate result, which is obtained under the condition of
large $N$. An important question is then how frequently we should perform
the tests. After each period of time, with some probability we get a wrong
state. This probability $P_{err}$ is measured by the norm of the amplitude
of the state outside the encoding space. Suppose the reservoir is initially
in thermal equilibrium. Following Eq. (5), $P_{err}$ can be expressed as $%
P_{err}=\frac \delta {N^2}$, where 
\begin{equation}
\label{9}
\begin{array}{c}
\delta =
\stackrel{2}{\stackunder{l=1}{\sum }}\int d\omega \left[ \left| g_{\omega
l}\right| ^2T_0^2\coth \left( \frac{\hbar \omega }{2k_BT}\right) \right]
+4Re\left( c_{+}^{*}c_{-}\right) \int d\omega \left[ Re\left( g_{\omega
1}^{*}g_{\omega 2}\right) T_0^2\coth \left( \frac{\hbar \omega }{2k_BT}%
\right) \right]  \\  \\ 
\leq \int d\omega \left[ \left( \left| g_{\omega 1}\right| +\left| g_{\omega
2}\right| \right) ^2T_0^2\coth \left( \frac{\hbar \omega }{2k_BT}\right)
\right] .
\end{array}
\end{equation}
After the time $T_0$, the accumulated error rate $P_{tot}$ is estimated by 
\begin{equation}
\label{10}P_{tot}\approx NP_{err}=\frac \delta N.
\end{equation}
Therefore, the scheme successfully prevents errors only when $N>>\delta $,
where the magnitude of $\delta $ is determined by the damping coefficients $%
\left| g_{\omega l}\right| ^2$, the evolution time $T_0$, and the
temperature $T$ of the reservoir. Eq. (10) suggests, a larger $N$, a smaller
accumulated error rate. However, this is not the case in practice, since we
need also consider the unavoidable noise introduced by the frequent tests.
Suppose $\gamma $ is the additional error rate introduced by the test after
each period of time. We assume that $\gamma $ satisfies $\gamma <<\delta $.
With this realistic consideration, the accumulate error rate is rewritten as 
\begin{equation}
\label{11}P_{tot}\approx \frac \delta N+N\gamma \geq 2\sqrt{\delta \gamma }.
\end{equation}
The minimum error rate is achieved if $N$ equals $\sqrt{\frac \delta \gamma }
$, which is the optimal value of the times of the tests. The error rate $%
\gamma $ introduced by every time test should be very small so that $2\sqrt{%
\delta \gamma }<<1$. This serves as the working condition of our scheme.

The above scheme can be extended to include multiple qubits. If we have $2L$
dissipative qubits, of course they can be exploited to protect $L$ qubits
information by encoding and detecting every two qubits in the way described
above. The efficiency $\eta $ of the scheme is $\frac 12$. This efficiency
can be further raised. For the two-qubit circumstance, we notice that the
encoding space is an eigenspace of the operator $B_1+B_2$. Similarly, for
the $2L$-qubit circumstance, the eigenspace of the operator $B_1+B_2+\cdots
+B_{2L}$ can also be used as the encoding space. An arbitrary state in this
eigenspace is mapped into another eigenspace by each of the operator $A_l$.
So the first-order error caused by the coupling to the reservoir can be
detected and prevented by measuring the observable $B_1+B_2+\cdots +B_{2L}$.
The largest eigenspace of the operator $B_1+B_2+\cdots +B_{2L}$ has a
dimension $\left( 
\begin{array}{c}
2L \\ 
L
\end{array}
\right) $, with the eigenvalue $0$. The maximum efficiency $\eta _{\max }$
thus attains 
\begin{equation}
\label{12}\eta _{\max }=\frac 1{2L}\log _2\left( 
\begin{array}{c}
2L \\ 
L
\end{array}
\right) \approx 1-\frac 1{4L}\log _2\left( \pi L\right) ,
\end{equation}
where the approximation is taken under the condition of large $L$. The
efficiency $\eta _{\max }$ is near to $1$ for large $L$.

We have shown that two qubits subject to noise described by the Hamiltonian
(1) are enough for protecting one qubit information. The Hamiltonian (1) is
quite general. Its different special case yields the coupling for amplitude
damping or for phase damping. Our scheme works whether the qubits couple
independently to separate environments or cooperatively to the same
environment. However, it is important to examine the type of the noise
beyond our description. The scheme relies on the Zeno effect, so it can deal
only with ''slow'' noise. The error should not accumulate too quickly [16].
The quantum Zeno effect dose not take place when the time interval between
the tests is larger than the characteristic time for which the exponential
decay approximation is applicable, therefore noise can be prevented only
when error occurs at a sub-exponential rate [23]. These limitations are also
suffered by other error-prevention schemes based on the Zeno effect [20,23].

The most favorable feature of the present scheme is that it prevents error
by costing few computing resources. This feature is remarkable since the
quantum computing resources are very stringent [28-30]. The proposed error
prevention code is very simple, so it has a good chance to be implemented in
experiments or to be used in numerical simulations of the robustness of
quantum information.\\

{\bf Acknowledgment}

This project was supported by the National Nature Science Foundation of
China.

\end{document}